\documentclass[prb,twocolumn,showpacs,preprintnumbers,amsmath,amssymb]{revtex4-1}

\usepackage{graphicx}
\usepackage{dcolumn}
\usepackage{bm}
\usepackage{amsmath}

\usepackage[dvips]{color}
\usepackage{ulem}

\begin{document}
\title
{
Crossover between two different Kondo couplings
in side-coupled double quantum dots
}

\author{Yoichi Tanaka$^1$}
\email{e-mail address: yoichiwaka@gmail.com}

\author{Norio Kawakami$^2$, and Akira Oguri$^3$}

\affiliation{
$^{1}$Condensed Matter Theory Laboratory, RIKEN, Saitama 351-0198, Japan\\
$^{2}$Department of Physics, Kyoto University, Kyoto 606-8502, Japan\\
$^{3}$Department of Physics, Osaka City University, Osaka 558-8585, Japan
}%

\date{\today}

\begin{abstract}
We study the Kondo effect in side-coupled double quantum dots 
with particular focus on the crossover between two distinct singlet 
ground states, using the numerical renormalization group.
The crossover occurs as the quantized energy level 
of the embedded dot, which is connected directly to the leads, is varied. 
In the parameter region where the embedded dot becomes 
almost empty or doubly occupied, the local moment emerging 
in the other dot at the side of the path for the current 
is screened via a superexchange process  
by the conduction electrons tunneling through the embedded dot. 
In contrast, in the other region where
the embedded dot is occupied by a single electron,
 the local moment emerges also in the embedded dot, and forms 
a singlet bond with the moment in the side dot.
Furthermore, we derive two different Kondo Hamiltonians for these limits
carrying out the Schrieffer-Wolff transformation, 
and show that they describe the essential feature 
of the screening for each case.

\end{abstract}


\maketitle

\begin{figure}[b]
\vspace{-1.5mm}
\includegraphics[scale=0.5]{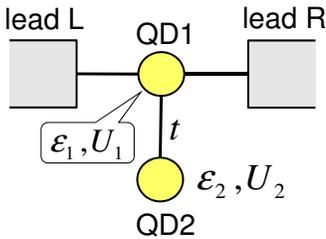}
\caption{
(Color online) 
Side-coupled double quantum dots.
$\varepsilon_{1(2)}$ and $U_{1(2)}$ are the energy level and 
the Coulomb interaction at the QD1 and QD2, respectively.
$t$ denotes the interdot coupling.
}
\label{modelTDQD}
\end{figure}

\section{Introduction}
The Kondo effect is a prototypical many-body phenomenon 
 that is caused by the interaction
between a localized spin and conduction electrons.
Since the Kondo effect was observed in a quantum dot (QD) 
system,\cite{Gold,Cronen} effects of electron correlation 
on quantum transport have attracted much attention.
Moreover, the recent experimental advancement enables one 
to examine the Kondo physics in a variety of systems, 
such as an Aharonov-Bohm ring with a QD and double quantum dots (DQD).
In these systems multiple paths for electron propagation 
also affect the tunneling currents, 
and give rise to a Fano-type asymmetry in the line shapes of conductance.

A side-coupled DQD system with a T-shape configuration, as
 shown in Fig.\ \ref{modelTDQD}, is a typical system in which 
the interplay between the Kondo effect and the interference 
effect occurs.
For this type of DQD, a number of  theoretical studies have been carried out
so far.
\cite{Kim,Taka,Corna,YT:2005,Luis,Zitko,Karr,Rams,Chung1,Chung2,Zitko2,
Maruyama2004,Serido}
It also becomes experimentally possible to fabricate this kind 
of geometry.\cite{Sasaki}
In the side-coupled DQD system shown in Fig.~\ref{modelTDQD},
one of the dots (QD2), which is referred to as a {\it side dot}, 
 has no direct coupling to the leads, 
but is coupled to the {\it embedded dot} (QD1).
Because of this unique geometry with the multiple paths, 
intriguing phenomena can occur in this system.
For example,  a two-stage Kondo effect can occur for the small 
interdot coupling $t$, where each of the dots 
is occupied by a single electron.
\cite{Corna,Zitko,Zitko2,Rams,Karr,Chung1,Chung2}
 In this case, the local moment in the QD1 is screened first by 
the conduction electrons at higher temperature, and   
then the moment in the QD2 is screened at lower temperature 
to form a singlet ground state. 
Most of the preceding studies of a side-coupled DQD system
focused on the two-stage Kondo effect taking place in this situation. 
In contrast, for large $t$, the two adjacent local moments
 screen each other to form a molecular-type singlet.

The gate voltage that is applied to the 
 dots can further change the charge and spin distributions in the DQD,
and evolve the system toward the mixed-valence regime.
Specifically, as the energy level $\varepsilon_1$ of the QD1 is varied,
another typical singlet state appears in the parameter region  
where the QD1 becomes almost empty or doubly occupied. 
It is a singlet bond between the local moment at QD2 and 
the conduction electrons, and is formed by a superexchange mechanism. 
Some numerical indications that this type of singlet state is formed  
were seen in data of previous works.\cite{Maruyama2004,Zitko}
Maruyama {\it et al.\/} obtained 
an asymmetric conductance peak of the Fano shape for finite $\varepsilon_1$ 
in the Kondo regime.\cite{Maruyama2004}
However, their study is focused mainly on the transport properties.
\v{Z}itko {\it et al.\/} \cite{Zitko} examined a similar situation, 
and showed that the Kondo temperature becomes small in this case.
The precise features of the singlet state, however, 
were not fully examined.
Some groups have studied the gate-voltage dependence of the conductance
 where the energy levels of the two dots are moved simultaneously.
\cite{Corna,Zitko,Karr}
However, it has still not been clarified in detail 
how the singlet ground state evolves
across the crossover region between 
the singlet state due to the two-stage Kondo effect
and the one formed by the superexchange mechanism 
between the QD2 and the conduction electrons.

In this paper, we re-examine a side-coupled DQD system shown 
in Fig.~\ref{modelTDQD}, and study how the Kondo 
 singlet bond is deformed as 
the energy level $\varepsilon_1$ in the QD1 is changed.
We find that as $\varepsilon_1$ moves away from the Fermi energy, 
the electrons at the QD1 cannot contribute to the screening
of the local moment at the QD2, and the conduction electrons tunneling
into the QD2 virtually via $\varepsilon_1$ screen the local moment.
This electron tunneling process is similar to the superexchange
 mechanism seen in transition metal oxides such as MnO and CuO.
\cite{Kramers,Anderson}
In the present case, the Kondo singlet bond is formed between 
the QD2 and the leads, mediated by the quantized level
 $\varepsilon_1$ at the QD1.
This mechanism is quite different from the Kondo screening 
in the case of $\varepsilon_1 \simeq 0$  
where a singlet bond between the QD1 and QD2 plays a dominant role.
\cite{Kim,Taka,Corna,YT:2005,Luis,Zitko,Karr,Rams,Chung1,Chung2,Zitko2,
Maruyama2004,Serido}
Therefore, the gate voltage applied to the QD1  
deforms the Kondo cloud, and it can be probed through 
the variation in the phase shift of the DQD.
In order to clarify these features, we calculate the phase shift
using the numerical renormalization group.
Furthermore, we calculate the spin susceptibility to obtain the 
Kondo temperature, which shows good agreement with the one 
obtained from an effective Kondo Hamiltonian 
we have derived in this work.
In the presence of the Coulomb interaction $U_1$ at the QD1, 
we also find that the two-stage Kondo screening changes to 
a single-stage process as $\varepsilon_1$ moves away from 
the electron-hole symmetric point $\varepsilon_1 \simeq -U_1/2$.

This paper is organized as follows.
In Sec.\ \ref{sec:model}, we give the Hamiltonian of our system. 
In Sec.\ \ref{sec:effH}, we derive the effective Hamiltonian
using the perturbation theory in the tunneling matrix elements,
which is identical to that derived from 
the Schrieffer-Wolff transformation.\cite{SW}
In Sec.\ \ref{sec:result}, we show the numerical results, 
and discuss how the Kondo singlet state evolves
as $\varepsilon_1$ varies.
A summary and discussions are given in Sec.\ \ref{sec:summary}.

\section{Model} \label{sec:model}

The Hamiltonian of a side-coupled DQD system 
shown in Fig.\ \ref{modelTDQD} reads
\begin{align} 
\,H = H_{QD1} + H_{QD2} + H_{\rm int} + 
\!\!\!
\sum_{\nu \in \{L,R \}} 
\!\!\!
(H_{\nu} + H_{T,\nu}),  
\label{Hamitotal}
\end{align}
where
\begin{align}
&H_{QDi}
=\varepsilon_{i} \sum_{\sigma} n_{i,\sigma} 
+ U_i n_{i,\uparrow}n_{i,\downarrow} , 
\nonumber\\
&
H_{\rm int}=
t\,\sum_{\sigma}\left(d_{1\sigma}^{\dag}d_{2\sigma}^{}+\textrm{H.c.}\right),
\,\,\,
H_{\nu}
=
\sum_{k,\sigma}\varepsilon _{k}
c_{\nu,k\sigma}^\dag c_{\nu,k\sigma}^{}  ,
\nonumber\\
&H_{T,\nu} = \sum_{k,\sigma} \frac{V_{\nu}}{\sqrt{\mathcal{N}}}
\left(c_{\nu,k\sigma}^\dag d_{1\sigma}^{} + \textrm{H.c.} \right),
 \quad \ \ \nu=L, R.
\label{Hamipart}
\end{align}
$H_{QDi}$ describes the QD1 for $i=1$,
and the QD2 for $i=2$,
$\varepsilon_{i}$ the energy level, 
$U_{i}$ the Coulomb interaction, and 
$n_{i,\sigma}=d^{\dag}_{i\sigma}d^{}_{i\sigma}$.
$H_{\rm int}$ denotes the interdot coupling
with the hopping matrix element $t$.
$H_{L/R}$ describes the normal lead of the left/right side. 
$V_{L/R}$  is the tunneling matrix element between the QD1 and 
the left/right lead. 
We assume that $\Gamma_{L/R}(\varepsilon) \equiv 
\pi V_{L/R}^{2} \sum_k \delta(\varepsilon-\varepsilon_{k})/\mathcal{N}$ 
is a constant independent of the energy $\varepsilon$, 
where $\mathcal{N}$ is the number of the states in each lead.

For convenience of the following discussions,
we apply a unitary transformation 
to  the leads, using the inversion symmetry, 
\begin{eqnarray}
&& \!\!\!\!\!\!\!\!\!\!\!\!
s_{k\sigma}=
\frac{V_L c_{L,k\sigma}+V_R c_{R,k\sigma}}{V_s}, \,
a_{k\sigma}=
\frac{V_L c_{L,k\sigma}-V_R c_{R,k\sigma}}{V_s},
\nonumber\\
&& \qquad\qquad V_s= \sqrt{{V_L}^2+{V_R}^2}.
\label{ope-sa}
\end{eqnarray}
Then, the Hamiltonian for the leads,
 $\sum_{\nu \in \{L,R \}} (H_{\nu} + H_{T,\nu})$,
can be rewritten as 
\begin{eqnarray}
\sum_{\nu \in \{L,R \}} 
\left( H_{\nu} + H_{T,\nu} \right)
&=&
H_s + H_a + H_{T,s} ,  
\label{HsaTs}
\end{eqnarray}
with
\begin{eqnarray}
&&
H_s = \sum_{k,\sigma}\varepsilon _{k}
s_{k\sigma}^\dag s_{k\sigma}^{},
\quad
H_a = \sum_{k,\sigma}\varepsilon _{k}
a_{k\sigma}^\dag a_{k\sigma}^{},
\label{HsHa}
\\
&&
H_{T,s} =
\sum_{k,\sigma} \frac{V_{s}}{\sqrt{\mathcal{N}}}
\left(s_{k\sigma}^\dag d_{1\sigma}^{} + \textrm{H.c.} \right).
\label{HTs}
\end{eqnarray}
Note that the operator $s_{k\sigma}$ couples to the QD1, 
while the operator $a_{k\sigma}$ is decoupled.
Therefore, we can map the original model given by Eq.\ \eqref{Hamitotal}
to the two-impurity Anderson model (TIAM),  
\begin{align} 
\mathcal{H}_{\rm TIAM} = H_{QD1} + H_{QD2} + H_{\rm int} + 
H_{s} + H_{T,s} .  
\label{Hami-TIAM}
\end{align}
Using this model, we will discuss how the singlet state due to 
the Kondo effect changes as 
the energy level $\varepsilon_1$ at the QD1 varies.

\section{Kondo Hamiltonians}
\label{sec:effH}

In this section, we derive the effective Hamiltonians 
in order to study the Kondo behavior in the two opposite cases,   
$\varepsilon_1 \simeq 0$ and large $\varepsilon_1$.
For this purpose, we use the perturbation theory in the tunneling 
matrix elements $t$ and $V_S$, which is equivalent to 
the Schrieffer-Wolff transformation.\cite{SW}
We assume that the Coulomb interaction at the 
QD1 is zero, $U_1=0$, in this section 
in order to focus on the effects of $\varepsilon_1$.
We also consider the effects of $U_1$ in Sec.\ \ref{sec:result}B
and the Appendix.

\begin{figure}[h]
\begin{center}
\includegraphics[scale=0.48]{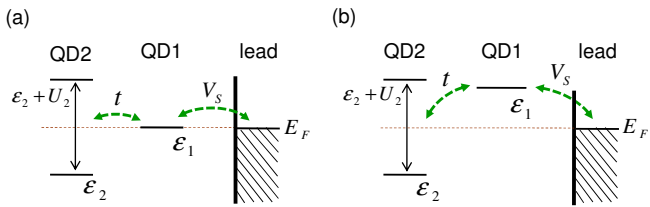}
\end{center}
\caption{
(Color online) 
Schematic energy diagram of a side-coupled DQD.
(a) $\varepsilon_{1} \simeq 0$, and
(b) $\varepsilon_{1} >0$ such that 
$\langle n_{1,\sigma} \rangle \simeq 0$ is satisfied.
$E_F$ denotes the Fermi energy of the leads.
}
\label{sketchTDQD}
\end{figure}

Figure \ref{sketchTDQD} shows a schematic energy diagram of a side-coupled DQD 
 for $\varepsilon_{2} + U_{2}/2 \simeq 0$.
Let us first consider the case, shown in Fig.\ \ref{sketchTDQD}(a),
where the energy level at the QD1 is located around the Fermi energy 
$\varepsilon_{1} \simeq 0$. 
When the Coulomb interaction $U_2$ is much larger than
the interdot coupling $t$, the local spin moment arises at the QD2.
Carrying out the perturbation expansion 
in the tunneling matrix element $t$,
we obtain the Kondo Hamiltonian of the form 
\begin{align} 
\mathcal{H}_{\rm K}^{\rm (a)}
= 
J_{12}\sum_{\sigma, \sigma'}
d^{\dag}_{1\sigma} \frac{\vec{\tau}_{\sigma \sigma'}}{2}
 d^{}_{1\sigma'} \cdot \vec{S}_2 + H_{QD1} + H_{s} + H_{T,s},
\label{Hami-Ka}
\end{align}
where
\begin{align} 
J_{12} &= 
2 t^2 \left\{
-\frac{1}{\varepsilon_2-\varepsilon_1}
+\frac{1}{(\varepsilon_2+U_2)-\varepsilon_1}
\right\}
\nonumber\\
&\xrightarrow{\varepsilon_1=0,\, \varepsilon_2 = -U_2/2}
\frac{8t^2}{U_2}.
\label{J-12}
\end{align}
$\vec{\tau}_{\sigma \sigma'} = (\tau_{\sigma \sigma'}^x, \tau_{\sigma \sigma'}^y, \tau_{\sigma \sigma'}^z)$ 
is the vector representation of the Pauli matrix, and $\vec{S}_2$ is 
the spin operator of the local spin at the QD2.
In the electron-hole symmetric case, 
$\varepsilon_1=0$ and $\varepsilon_2 = -U_2/2$, 
the exchange coupling is given by $J_{12} =8t^2/U_2$. 
This type of exchange coupling appears in the usual Kondo model,
 and is referred to as the ``interdot exchange (IE)" coupling in the following,
 in order to distinguish it from 
the superexchange coupling discussed later.
In the case where $J_{12}$ is larger than $\Gamma$ ($\equiv \Gamma_L+\Gamma_R$), the local moments at the QD1 and QD2 form 
a molecular-type singlet bond.
In the opposite case $J_{12} < \Gamma$ where 
the QD1 is coupled more strongly to the leads,
the spectral weight of the electron at the QD1 is broadened, 
and the local moment $\vec{S}_2$ at the QD2 is 
 screened by the electrons with this broadened density of states.
In both of these cases, the singlet state 
is formed mainly between the moments in the QD1 and QD2. 
Similar considerations also make sense 
for finite $U_1$,\cite{Corna,Zitko} just by replacing $\Gamma$ with 
the Kondo temperature $T_K^{QD1}$ for the QD1.  
Namely, the two-stage Kondo effect occurs 
for $J_{12}<T_K^{QD1}$, whereas a singlet bond becomes 
the molecular-type one for $J_{12}>T_K^{QD1}$.

Next, we consider the situation shown in Fig.\ \ref{sketchTDQD}(b),
where the energy level at the QD1 is away from the Fermi energy.
The Kondo effect in this situation is a main focus of this paper.
In this case, the QD1 is almost empty (doubly occupied)
for $\varepsilon_1>0$ ($\varepsilon_1<0$),
and the spin degree of freedom disappears at the QD1.
Thus, the electrons at the QD1 can not contribute to the screening
of the local moment at the QD2, and
 $\varepsilon_1$ works as a potential barrier 
that disturbs charge transfer between the QD2 and the leads. 
Maruyama \textit{et al.} \cite{Maruyama2004}
and \v{Z}itko \textit{et al.} \cite{Zitko} also 
 considered the $|\varepsilon_1| \to \infty$ limit.
However, the precise features of the screening process
due to the superexchange mechanism were not examined in detail.
A similar situation to Fig.\ \ref{sketchTDQD}(b) also arises
in transition metal oxides such as MnO and CuO, 
where the superexchange interaction describes the coupling 
between the local spins in magnetic ions 
mediated by nonmagnetic oxygen anions.
\cite{Kramers,Anderson}

In order to clarify the Kondo screening in 
the situation shown in Fig.\ \ref{sketchTDQD}(b), 
we derive the effective Hamiltonian from
the  perturbation expansion 
with respect to the tunneling elements $t$ and $V_{s}$ 
to the fourth order (see the Appendix). 
The result can be expressed in the form
\begin{align} 
\mathcal{H}_{\rm K}^{\rm (b)} = 
J_{\rm SE} \sum_{k, k'} \sum_{\sigma, \sigma'}
s^{\dag}_{k\sigma} \frac{\vec{\tau}_{\sigma \sigma'}}{2}
 s^{}_{k'\sigma'} \cdot \vec{S}_2
 + H_{s},
\label{Hami-Kb}
\end{align}
where
\begin{align} 
J_{\rm SE} =
2 \left( \frac{V_s t}{\varepsilon_1} \right)^2 
\!\!
\left(
-\frac{1}{\varepsilon_2}
+\frac{1}{\varepsilon_2+U_2}
\right)
\xrightarrow{\varepsilon_2 = -\frac{U_2}{2}}
\frac{8}{U_2}
\left( \frac{V_s t}{\varepsilon_1} \right)^2_.
\label{J-SE}
\end{align}
The coupling constant $J_{\rm SE}$ 
depends on the energy level $\varepsilon_{1}$,
 which is caused by a virtual process with a single electron 
 passing through the QD1.
This term appears as the $V_s^2 t^2 $-type contribution  
in the fourth order perturbation expansion 
with respect to tunneling elements.
The screening of the local moment $\vec{S}_2$ 
is achieved for large $\varepsilon_1$ 
by the conduction electrons tunneling virtually through the QD1.  
This screening mechanism is essentially the same as the 
one due to the superexchange interaction mentioned above,
and the singlet bond becomes long compared to 
that in the case of $\varepsilon_1 \simeq 0$. 
Therefore, $J_{\rm SE}$ is referred to as
the ``superexchange (SE)" Kondo coupling in the following.
Note that a similar screening occurs 
also for negative $\varepsilon_{1} (<0)$, 
although Fig.\ \ref{sketchTDQD}(b) describes 
only the situation for positive $\varepsilon_{1} (>0)$.
For negative large $\varepsilon_1$, the QD1 is almost doubly occupied 
and the effective Hamiltonian takes the same form
$\mathcal{H}_{\rm K}^{\rm (b)}$ in Eq.\ \eqref{Hami-Kb}.

\section{NRG Results} \label{sec:result}

\subsection{Crossover between the IE and SE Kondo screenings}

To confirm the above discussions more precisely, 
we calculate the phase shift, the average number of electrons 
in each of the dots, and the spin susceptibility
for the two-impurity Anderson model $\mathcal{H}_{\rm TIAM}$
using the numerical renormalization group (NRG).\cite{Wilkins}
We first show the numerical results of the phase shift due to the DQD,
which is helpful to clarify the formation of the singlet state 
because the phase shift reflects an electron scattering at the DQD. 
 From the phase shift $\varphi$, we can also deduce 
the total number of electrons 
$N_{DQD} \equiv \sum_{\sigma} \langle n_{1,\sigma} + n_{2,\sigma}\rangle$ 
in the DQD, using the Friedel sum rule \cite{Langer, Sum:Tanaka}
\begin{align}
N_{DQD} \,=\, \frac{2}{\pi}\varphi  \;. 
\label{eq:phase_shift}
\end{align}
\begin{figure}[b]
\begin{center}
\includegraphics[scale=0.6]{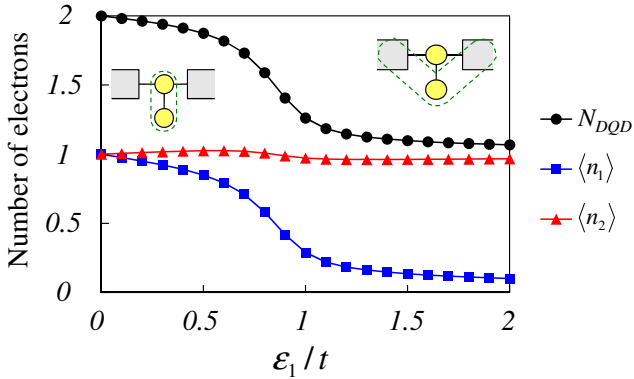}
\end{center}
\caption{(Color online)
Total number of electrons in the DQD, $N_{DQD}$,
and each number of the electrons at the dots.
We set $U_1=0$, $U_2/t=6$, $\varepsilon_2=-U_2/2$,
and $\Gamma_L/t=\Gamma_R/t=0.1$.
The inserted figures illustrate a formation of
a dominant singlet bond which is described by the (green) dashed line.
}
\label{NP}
\end{figure}
Figure \ref{NP} shows $N_{DQD}$ as a function of $\varepsilon_1/t$,
and the number of electrons in each of the dots,
$\langle n_{1(2)} \rangle = \sum_{\sigma} \langle n_{1(2),\sigma} \rangle$.
Let us first look at a region around $\varepsilon_1=0$.
In this region, namely $N_{DQD} \simeq 2$,
 both of the dots are nearly half-filled
($\langle n_{1(2)} \rangle \simeq 1$), and 
the phase shift takes the value of 
 $\varphi \simeq \pi$ from Eq.\ \eqref{eq:phase_shift}.
 The singlet state is formed dominantly
inside the DQD, and thereby the conduction electrons at the leads
are not scattered by the local spin at the dots. 
As $\varepsilon_1$ increases,  $N_{DQD}$ shows a sharp drop around 
$\varepsilon_1/t=0.8$ and approaches $N_{DQD} \simeq 1$.
 From Eq.\ \eqref{eq:phase_shift}, we see that the phase shift $\varphi$
also changes from $\pi$ to $\pi/2$. 
In the region around $\varepsilon_1/t=2.0$, $n_{1}$ goes to zero while $n_{2}$
almost remains unchanged. 
It indicates that the local spin appears only at the QD2 
for large $\varepsilon_1$, and the SE Kondo coupling can be 
described by the Kondo Hamiltonian given in Eq.\ \eqref{Hami-Kb}. 
Therefore, the kink behavior of the phase shift with  
the height $\pi/2$, seen in Fig.\ \ref{NP}, signifies 
the crossover between the IE Kondo screening described by 
the Hamiltonian $\mathcal{H}_{\rm K}^{\rm (a)}$ with $J_{12}$
and the SE one described by $\mathcal{H}_{\rm K}^{\rm (b)}$ 
 with $J_{\rm SE}$.

\begin{figure}[t]
\begin{center}
\includegraphics[scale=0.64]{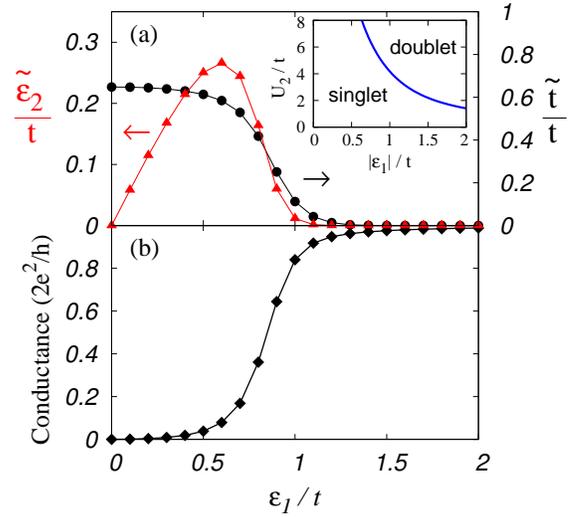}
\end{center}
\caption{(Color online)
(a)
Renormalized parameters $\widetilde{\varepsilon}_{2}$, $\widetilde{t}$
 and
(b) conductance at zero temperature as a function of $\varepsilon_{1}/t$.
The parameters of the system are the same as in Fig. \ref{NP}.
Inset of (a):
Phase boundary between singlet and doublet ground states 
for the isolated DQD system ($\Gamma_{L/R}=0$),
where we set $U_1=0$ and $\varepsilon_2=-U_2/2$.
}
\label{reCon}
\end{figure}

In order to discuss the change of the low-energy states 
for the energy level $\varepsilon_1$ in more detail,
it is helpful to use the fixed-point Hamiltonian in terms of
the renormalized parameters \cite{Hewson:qp}
\begin{align}
& \widetilde{H}_{qp}^{(0)}  =   
\widetilde{\varepsilon}_{2} n_2
+\varepsilon_{1} n_1
+\widetilde{t}\, \sum_{\sigma}\left(d_{1\sigma }^\dag d_{2\sigma}^{}
+\textrm{H.c.}\right)
\nonumber\\
& \qquad \quad 
+\sum_{\nu} (H_{\nu} + H_{T,\nu}) ,
\label{Hamiqp}
\end{align}
where 
\begin{align}
\widetilde{\varepsilon}_{2}  \equiv   
Z \,(\varepsilon_{2} + \Sigma_{2}(0)) , \quad 
\widetilde{t} \equiv  \sqrt{Z} \, t , 
\nonumber\\
Z \equiv   
\left(
1- \left.\!  
\frac{\partial \Sigma_{2}(\varepsilon)}{\partial \varepsilon}
\right|_{\varepsilon=0} 
\right)^{-1}.
\label{eq:Z_til} 
\end{align}
$\Sigma_{2}(\varepsilon)$ is the self energy due to the Coulomb interaction 
$U_2$. 
Using this fixed-point Hamiltonian,  
a unified analysis for the $\varepsilon_{1}$ dependence becomes possible.
 The crossover between the two opposite limits, at $\varepsilon_1 = 0$ 
and $\varepsilon_1 \to \infty$, 
can be described as a continuous change of 
the parameter values of the fixed-point Hamiltonian. 
We can calculate the renormalized parameters 
$\widetilde{\varepsilon}_{2}$ and $\widetilde{t}$ 
using the NRG.\cite{Hewson2}
The results are shown in Fig.\ \ref{reCon}(a).
It is noteworthy that both $\widetilde{\varepsilon}_{2}$ and 
$\widetilde{t}$ show a sharp decrease around $\varepsilon_{1}/t \simeq 0.8$, 
which indicates the crossover between the two different
singlet bonds, namely the one due to the IE coupling $J_{12}$ 
and the other due to the SE Kondo coupling $J_{\rm SE}$.
Furthermore, from these parameters we can deduce the phase shift 
$\varphi$ of the DQD and the conductance $G$ at zero temperature:
\cite{Corna}
\begin{eqnarray}
\varphi
&=&
\frac{\pi}{2}+{\rm tan}^{-1}\left(
\frac{\widetilde{t}^2-\widetilde{\varepsilon}_{2} \varepsilon_{1}}
{\widetilde{\varepsilon}_{2}\Gamma}
\right)
\, ,
\label{eq:phase_shift_para}
\\
 G&=&\frac{2e^2}{h} \sin^2 \varphi
=\frac{2e^2}{h} 
\left\{
1+\left( \frac{\widetilde{t}^2-\widetilde{\varepsilon}_{2} \varepsilon_{1}}
{\widetilde{\varepsilon}_{2}\Gamma}
 \right)^2
\right\}^{-1}_,
\label{condu}
\end{eqnarray}
where $\Gamma \equiv \Gamma_L+\Gamma_R$. 
Note that this expression for $G$ is exact 
at zero temperature for the symmetric coupling $\Gamma_L=\Gamma_R$, 
and can be obtained, for instance,  by using
the Meir-Wingreen formula\cite{MW} for 
the Hamiltonian in Eq.\ \eqref{Hamitotal}. 
The phase shift $\varphi$ and the conductance $G$ can be deduced from  
 the exact NRG results for the renormalized parameters. 
Figure \ref{reCon}(b) shows the result of the conductance
 as a function of $\varepsilon_{1}$ for $\varepsilon_2=-U_2/2$.
We see that the conductance shows an upturn around 
$\varepsilon_{1}/t \simeq 0.8$, and at this 
value the crossover between the ground state due to the IE coupling 
and that due to the SE Kondo coupling occurs.
The behavior of the conductance in Fig. \ref{reCon}(b) can also 
be explained in terms of the Fano-Kondo effect.
This is because the energy level $\varepsilon_{1}$ at the QD1 
 varies the asymmetric parameter $q$ for a Fano line shape,
as discussed by Maruyama {\it et al.} \cite{Maruyama2004} and 
\v{Z}itko.\cite{Zitko2} 
Indeed, the conductance decreases 
at $\varepsilon_{1}/t \simeq 0$, where $q\simeq 0$, 
due to the destructive interference effect
 while the conductance approaches $2e^2/h$ in the limit 
of $\varepsilon_{1}/t \to \infty$ where $q\to \infty$. 

The nature of the crossover can also be related 
to a level crossing taking place 
in a \textit{molecule} limit $\Gamma_{L/R}=0$, 
where the QD1 is decoupled from the lead.
In this limit the isolated DQD is described 
by a Hamiltonian $H_{QD1} + H_{QD2} + H_{\rm int}$,
and the ground state of the \textit{molecule} becomes a singlet or doublet, 
depending on the value of $|\varepsilon_1|/t$ and $U_2/t$.
As shown in the inset of Fig.\ \ref{reCon}(a) for $U_{1}=0$,
the ground state is a spin singlet  
if either $|\varepsilon_1|/t$ or $U_2/t$ is small. 
In the opposite case, a spin doublet becomes the ground state.
Note that in the doublet region, nearly one electron occupies the QD2, 
whereas the QD1 is almost empty or doubly occupied.
Thus, the local moment emerges mainly at the QD2.
We see in the phase diagram in Fig.\ \ref{reCon}(a) that
the transition takes place in this \textit{molecule} limit
at $|\varepsilon_1|/t \simeq 0.8$ for $U_2/t=6$,
and it agrees well with the position where 
$\widetilde{\varepsilon}_{2}$ and $\widetilde{t}$ show a
sharp decrease.
For finite $\Gamma_{L/R}$, the conduction electrons can tunnel 
from the lead to the QD2 via the QD1.
However, the electrons at the QD1 cannot contribute to the screening of 
the moment at the QD2, because the QD1 is almost empty or doubly occupied, 
and has no local spin moment.
Then, the local spin at the QD2 is screened by the 
conduction electrons from the leads over the QD1, which is the 
SE Kondo screening discussed above. 
Therefore, using the phase diagram in Fig.\ \ref{reCon}(a),
we can estimate the value  $|\varepsilon_1|/t$ and $U_2/t$, 
 at which a crossover between two distinct singlet states occurs.

In order to estimate the Kondo temperature due to the SE process,
we calculate the contribution of the QD2 to the impurity susceptibility,
 defined by
\begin{align} 
\chi_{2}
=
\frac{(g\mu_B)^2}{k_B T}  
\left( \langle S_z^2 \rangle - \langle S_z^2 \rangle_0 \right),
\label{eq:susQD2}
\end{align}
where $\langle S_z^2 \rangle$ ($\langle S_z^2 \rangle_0$) 
is the $z$ component of the total spin 
of the system with (without) the QD2.
\begin{figure}[h]
\begin{center}
\includegraphics[scale=0.65]{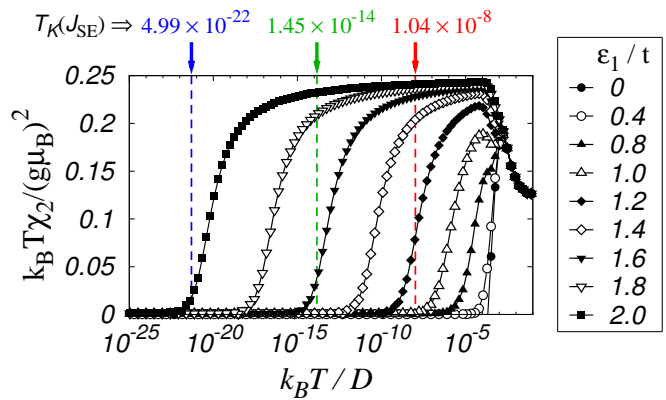}
\end{center}
\caption{
(Color online)  
Plots of $k_B T \chi_{2}/(g\mu_B)^2$ vs $k_B T/D$ for several values of
$\varepsilon_{1}/t$.
We set $U_2/t=6, \varepsilon_{2}=-U_2/2$, $\Gamma_L/t=\Gamma_R/t=0.1$,
 and $t/D=10^{-3}$.
The arrows with dashed lines indicate the 
Kondo temperature $T_K(J_{\rm SE})$ obtained from Eq.\ \eqref{TKfromJ}
for $\varepsilon_1/t=1.2$, $1.6$, and $2.0$ (see Table \ref{tab:table4}).
}
\label{susQD2}
\end{figure}
Figure \ref{susQD2} shows the results of $\chi_{2}$ for several 
values of $\varepsilon_1$.
We can estimate the Kondo temperature from the slope
of these plots.\cite{Wilkins}
In particular, for a large value of $\varepsilon_1$, 
we can compare the Kondo temperature estimated from $T\chi_{2}$ with 
that obtained from the 
Kondo Hamiltonian $\mathcal{H}_{\rm K}^{\rm (b)}$ 
with a formula\cite{Wilkins}
\begin{eqnarray} 
T_K(J_{\rm SE}) 
= D \sqrt{\rho J_{\rm SE}} \,\exp(-1/\rho J_{\rm SE}),
\label{TKfromJ}
\end{eqnarray}
where $D$ is the half-bandwidth of the leads and $\rho=1/2D$.
\begin{table}[h]
\caption{\label{tab:table4}
The Kondo temperatures $T_K(J_{\rm SE})$ from Eq.\ \eqref{TKfromJ}.}
\begin{ruledtabular}
\begin{tabular}{cccc}
$\varepsilon_1$/t&1.2&1.6&2.0\\
\hline \\[-2.5mm]
$T_K(J_{\rm SE})/D$ & $1.04\times 10^{-8}$ & $1.45\times 10^{-14}$&
$4.99\times 10^{-22}$\\
\end{tabular}
\end{ruledtabular}
\end{table}

The Kondo temperature $T_K(J_{\rm SE})$, 
which is obtained from Eq.\ \eqref{TKfromJ}, 
is listed in Table \ref{tab:table4}  
for several values of $\varepsilon_1/t$. 
Furthermore, these values of $T_K(J_{\rm SE})$ are 
indicated by the arrows with dashed lines in Fig.\ \ref{susQD2}. 
 We see that $T\chi_{2}$ obtained from the NRG
for $\varepsilon_1/t=1.2$, $1.6$, and $2.0$ 
 decreases rapidly showing a clear crossover to the Kondo regime
around the temperature indicated by the arrows.
This agreement demonstrates that the Kondo screening
for a large value of $\varepsilon_1$ is mainly owing to the SE one 
described by the Kondo Hamiltonian $\mathcal{H}_{\rm K}^{\rm (b)}$.

\subsection{Coulomb interaction $U_1$ at QD1}

So far, we have assumed that the QD1 is noninteracting, $U_1=0$.
In this subsection, 
we discuss the effects of the Coulomb interaction $U_1$ at the QD1 
on the energy scale of the Kondo screening.

\subsubsection{Finite $U_1$ for $\varepsilon_1>0$}

We first introduce the Coulomb interaction $U_1$ 
for positive $\varepsilon_1>0$.
\begin{figure}[h]
\begin{center}
\includegraphics[scale=0.65]{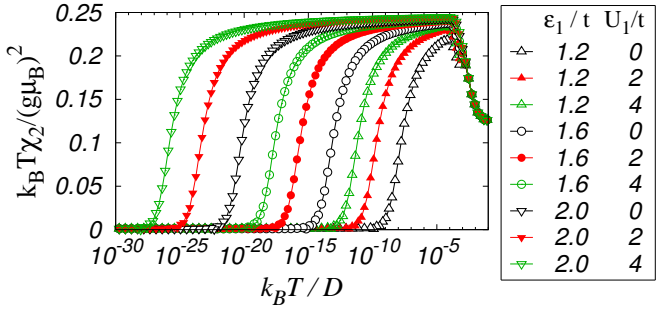}
\end{center}
\caption{
(Color online)  
Plots of $k_B T \chi_{2}/(g\mu_B)^2$ vs $k_B T/D$ for several values of
$\varepsilon_{1}/t$ and $U_1/t$, where we set 
$U_2/t=6, \varepsilon_{2}=-U_2/2$, $\Gamma_L/t=\Gamma_R/t=0.1$,
and $t/D=10^{-3}$.
}
\label{susQD2finiteU1}
\end{figure}
Figure \ref{susQD2finiteU1} shows the impurity susceptibility at the QD2,
$\chi_{2}$, for several values of $\varepsilon_1$ and $U_1$.
The other parameters are the same as those of Fig.\ \ref{susQD2}.
Since we set $\varepsilon_1$ in a way such that
there is almost no local spin moment at the QD1, 
the SE Kondo screening occurs in these examples.
In Fig.\ \ref{susQD2finiteU1}, we see that the crossover temperature, 
corresponding to the Kondo energy scale,  decreases as $U_1$ increases.
This is because the QD1 becomes almost empty 
for large positive $\varepsilon_1$ and large Coulomb repulsion $U_1$, 
and the virtual electron tunneling from the leads 
to QD2 is suppressed significantly.
Specifically, in the case where $\varepsilon_1$ 
is away from the Fermi energy ($\varepsilon_1>t, \Gamma$),
the SE Kondo coupling $J_{\rm SE}$  
between the QD2 and the leads given in Eq.\ \eqref{J-SE}
can be expressed in a more general form, 
by taking into account the effect of the Coulomb repulsion $U_1$, as
\begin{eqnarray} 
J_{\rm SE}
=
2 \!
\left( \frac{V_s t}{\varepsilon_1} \right)^{\! 2} 
\!\!
\left\{
\frac{\varepsilon_1^2 U_1- (\varepsilon_2-2\varepsilon_1)
(\varepsilon_1-\varepsilon_2)^2 }
{\varepsilon_2 (\varepsilon_2-2\varepsilon_1-U_1)
(\varepsilon_1-\varepsilon_2)^2}
+
\frac{1}{\varepsilon_2+U_2}
\right\}_.
\nonumber\\
\label{J-SE-U1}
\end{eqnarray}
This $J_{\rm SE}$ monotonically decreases with increasing $U_1$,
 so that the Kondo temperature $T_K(J_{\rm SE})$ defined by 
Eq.\ \eqref{TKfromJ} also becomes small as $U_1$ increases.
We can see the corresponding shift of the crossover temperature 
in Fig.\ \ref{susQD2finiteU1} for the two-impurity Anderson model
$\mathcal{H}_{\rm TIAM}$.

\subsubsection{
Crossover between the two-stage and single-stage Kondo screenings
}

Next we consider another case: the crossover between 
the two-stage Kondo screening and a single-stage Kondo screening.
It has been discussed previously that 
the two-stage Kondo effect can occur in the case where 
each of the two dots has a local moment and 
the Kondo temperature $T_{K}^{QD1}$ for the QD1 is larger 
than the exchange coupling $J_{12}$ between the dots.\cite{Corna,Zitko}
It takes place typically near the electron-hole symmetric point, 
where $\varepsilon_{1(2)} + U_{1(2)}/2 \simeq 0$. 
Therefore, as the energy level of the QD1 
moves away from the electron-hole symmetric 
point $\varepsilon_1 \simeq -U_1/2$, the SE Kondo screening
 can arise because the local spin moment at the QD1 disappears.
 Furthermore, the two-stage Kondo screening near the symmetric point 
changes to the single-stage Kondo screening.
This was also discussed partly by \v{Z}itko {\it et al.} 
They showed that the Kondo temperature rapidly drops 
as $\varepsilon_1$ moves away from the electron-hole symmetric point.
\cite{Zitko}
However, how the singlet ground state evolves in the crossover region
has not been clarified in detail.
In order to confirm the precise features of the Kondo screening, 
we calculate the susceptibility 
for the two dots as well as that for the QD2,
\begin{align} 
\chi_{\rm DQD}
=
\frac{(g\mu_B)^2}{k_B T}  
\left( \langle S_z^2 \rangle - \langle S_z^2 \rangle_{\rm lead} \right).
\label{eq:susDQD}
\end{align}
Here, $\langle S_z^2 \rangle$ is the $z$ component of the total spin
of the whole system including the DQD, and 
$\langle S_z^2 \rangle_{\rm lead}$ is the same quantity without the DQD.
Furthermore, we also calculate the entropy of the DQD defined by
\begin{align} 
S_{\rm DQD}
=
\frac{1}{k_B T}
\Bigl\{ (E-F) - (E_{\rm lead} -F_{\rm lead}) \Bigr\}.
\label{entro}
\end{align}
Here, $E={\rm Tr}[H e^{-H/(k_BT)}] 
/{\rm Tr}[e^{-H/(k_BT)}]$
and $F=-k_BT\,{\rm ln}{\rm Tr}[e^{-H/(k_BT)}]$
are the internal energy and free energy of the whole system consisting
 of the DQD and the leads, while $E_{\rm lead}$ and $F_{\rm lead}$
are those for the unconnected leads.

\begin{figure}[t]
\begin{center}
\includegraphics[scale=0.65]{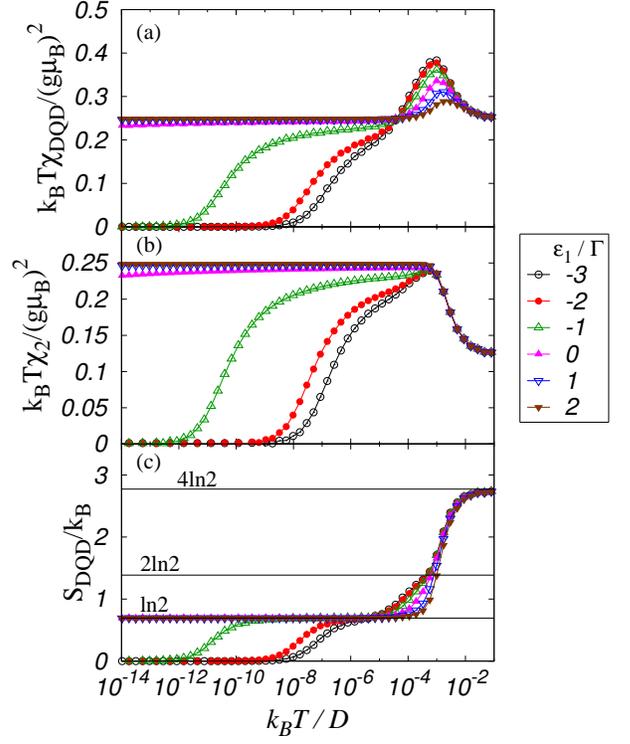}
\end{center}
\caption{
(Color online)
Spin susceptibility (a) $k_B T \chi_{\rm DQD}/(g\mu_B)^2$,
(b) $k_B T \chi_{\rm 2}/(g\mu_B)^2$, and
(c) entropy $S_{\rm DQD}/k_B$ as a function of $k_B T/D$
for several values of $\varepsilon_1/\Gamma$.
We set $t/\Gamma=0.3$, $U_1/\Gamma=U_2/\Gamma=6$, 
$\varepsilon_{2}/U_2=-0.5$, and $\Gamma/D=10^{-3}$.
}
\label{susentDQD}
\end{figure}
\begin{figure}[h]
\begin{center}
\includegraphics[scale=0.65]{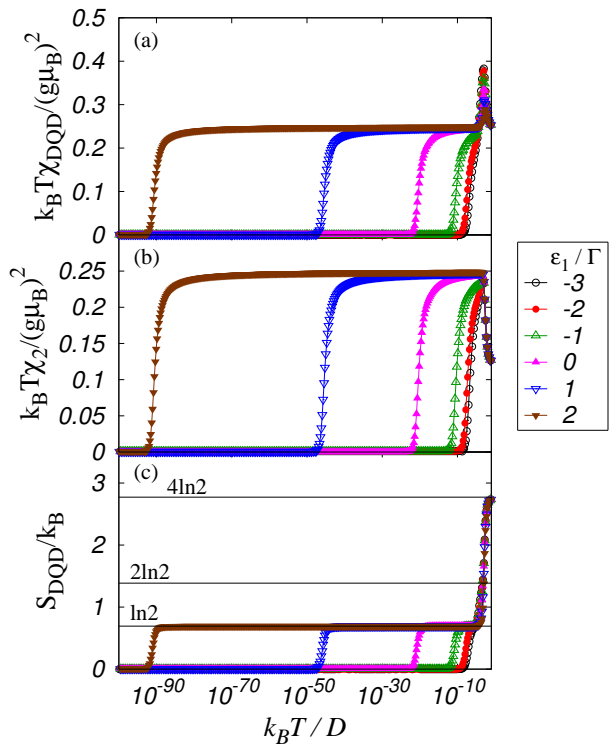}
\end{center}
\caption{
(Color online) 
The results presented in Fig.\ \ref{susentDQD} 
are replotted over a wide temperature range.
These results confirm that the moment is eventually  
screened in the cases of $\varepsilon_1/\Gamma=1$ and $2$ 
although the screening temperature becomes exponentially small.
Therefore, to observe the screening process due to the SE Kondo coupling
$J_{\rm SE}$ defined by Eq.\ \eqref{J-SE-U1}
 in a realistic temperature, $\varepsilon_1$ should be not so large. 
}
\label{susentDQD-lowerT}
\end{figure}

Figure \ref{susentDQD}(a) shows the spin susceptibility $\chi_{\rm DQD}$ 
 for several values of $\varepsilon_1/\Gamma$.
For $\varepsilon_1/\Gamma=-3$,
which corresponds to the electron-hole symmetric point $\varepsilon_1+U_1/2=0$,
$T \chi_{\rm DQD}$ shows a peak around $T/D \sim 10^{-3}$.
This peak indicates that each of the dots is occupied by a single electron, and the local moment is well developed. 
In this case, a two-stage screening occurs as temperature decreases:
the first stage can be seen at $T/D \sim 10^{-4}$, 
and the second one at $T/D \sim 10^{-7}$.
These two energy scales correspond to 
the Kondo temperature for the first stage $T_K^{\rm 1st}$
and that for the second stage $T_K^{\rm 2nd}$, respectively.
If the Coulomb interaction at the dots is much larger than 
the tunneling constants $\Gamma$ and $t$, 
then the peak of $T \chi_{\rm DQD}$ approaches $0.5$ 
in units of $(g\mu_\mathrm{B})^2$ and the structure
that emerges at $T_K^{\rm 1st}$ becomes clear.
For $\varepsilon_1/\Gamma=-2$ and $-1$,
which are still negative but closer to the Fermi energy,
the two-stage Kondo effect can be seen more clearly.   
The Kondo temperature $T_K^{\rm 2nd}$ 
for the second stage becomes lower: 
$T_K^{\rm 2nd}/D \sim 10^{-8}$ for $\varepsilon_1/\Gamma=-2$
and $T_K^{\rm 2nd}/D \sim 10^{-11}$ for $\varepsilon_1/\Gamma=-1$.
In Fig.\ \ref{susentDQD}(c), we see that 
the two-stage behavior can be observed more sharply 
in the temperature dependence of the entropy than 
that of the susceptibility.

As $\varepsilon_1$ crosses the Fermi energy and takes a positive value,
the peak of $T \chi_{\rm DQD}$ seen 
in Fig.\ \ref{susentDQD}(a) at $T/D \sim 10^{-3}$ is suppressed. 
This indicates that the local moment at the 
QD1 disappears as the QD1 becomes almost empty.
Then, the local moment at the QD2 is screened by the SE process 
via a single stage as the temperature approaches zero.
We see in Fig.\ \ref{susentDQD}(b) that
 $T \chi_{2}$ is almost constant (0.25) in a 
wide temperature range below $T/D \sim 10^{-3}$.
This indicates that the local spin moment at the QD2 remains 
almost free in this temperature region. 
Furthermore, the entropy for the DQD shown in  
Fig.\ \ref{susentDQD}(c) is locked 
at the value of ${\rm ln}2$ below $T/D \sim 10^{-4}$,
which is caused by the unscreened spin moment at the QD2.

This free moment must be screened eventually at low temperatures
by the SE Kondo coupling $J_{\rm SE}$
although the Coulomb interaction $U_1$ makes 
the Kondo temperature very small.
In order to see the low temperature region, 
 $\chi_{\rm DQD}$, $\chi_{2}$, and $S_{\rm DQD}$ are shown in
 a wide temperature range in Fig.\ \ref{susentDQD-lowerT}.
We see that $T\chi_{\rm DQD}$ for $\varepsilon_1/\Gamma=1$ 
and $\varepsilon_1/\Gamma=2$ shows the decrease at $T/D \sim 10^{-45}$ 
and $T/D \sim 10^{-90}$, respectively.
These results confirm clearly that 
the local moment at the QD2 is really screened 
although the energy scale for the SE Kondo screening
becomes small for large $U_1$ and $\varepsilon_1$.

\section{Summary and discussion} \label{sec:summary}
We have studied the Kondo effect in a side-coupled DQD system with
focus on how the Kondo singlet state changes by varying
the energy level at the embedded dot (QD1).
We have found that when the side dot (QD2) is in the Kondo regime,
two distinct singlet states appear; one is due to the IE coupling
 between the QD1 and the QD2, and the other is caused by the SE Kondo coupling
between the QD2 and the leads via the QD1.
In this sense, the latter is a different type of the singlet state 
from the former which has been studied in the side-coupled DQD systems 
so far.
In order to clarify the screening process,
we have obtained the effective Kondo Hamiltonians
using the perturbation expansion with respect 
to the tunneling matrix elements.
 From these Kondo Hamiltonians, we have shown that 
in the case where the QD1 is almost empty and doubly occupied
the screening is caused by a superexchange mechanism,
 and the singlet bond becomes long.
Moreover, we have calculated the phase shift and the conductance
using the NRG method, and have obtained the relation 
between the phase shift and the conductance.
We have found that the conductance is enhanced at the crossover region 
between the singlet ground state due to the IE coupling 
and that due to the SE Kondo coupling.
We have also calculated the local spin susceptibility
and have estimated the Kondo temperature,
which shows good agreement with that
obtained from the effective Kondo Hamiltonian.
Furthermore, we have demonstrated precisely how the two-stage Kondo screening 
changes to a single-stage process as the energy level at the QD1 
moves away from the electron-hole symmetric point.


In closing, we would like to make some comments on the SE Kondo screening. 
The scenario of the SE Kondo screening is not limited 
to our side-coupled DQD system, but is more generic for
 the Anderson model where the impurity spin 
and the conduction electrons are connected via the discrete energy level.
In usual cases of the Kondo problem,
a magnetic impurity is coupled directly to the conduction electrons,
for instance, a bulk system with a magnetic impurity.
On the other hand, in the two-impurity Anderson model we have considered,
 a local spin moment is coupled indirectly to the conduction electrons via 
a discrete energy level. The resulting screening process
 shows a unique feature, which has clearly been demonstrated
in this paper with the help of the effective Kondo Hamiltonians
and the NRG method.
The SE Kondo screening discussed in this paper
also appears in a side-coupled DQD
system coupled to normal and superconducting leads,
which we have previously studied.\cite{Tanaka:NTDQDS}
In this case, the superconducting proximity to the embedded dot
 quenches the local spin moment because this proximity
tends to make a singlet consisting of a linear combination
of the empty and doubly occupied states.
Thus, the superconducting proximity to the embedded dot
 plays the role of a potential barrier
between the side dot and the normal lead,
which can cause the SE Kondo screening.

Recently, the side-coupled DQD system has been fabricated 
in experiments.\cite{Sasaki}
We thus expect that in the near future, it may become possible to
observe the SE Kondo screening discussed in this paper, 
providing further interesting examples of correlation effects 
in the context of electron transport in nanoscale systems.


\begin{acknowledgments}
Y.T.\ was supported by the Special Postdoctoral Researchers Program of RIKEN.
N.K. is supported by JSPS FIRST-Program, the Grant-in-Aid for Scientific Research [Grant Nos. 21540359 and 20102008], and the Global COE Program ``The
Next Generation of Physics, Spun from Universality and Emergence" from
MEXT of Japan.
A.O.\ is supported by JSPS Grant-in-Aid for Scientific Research (C)
 (Grant No.\ 23540375).
\end{acknowledgments}

\appendix

\section*{Appendix: Derivation of the effective Kondo Hamiltonian} 
\label{sec:Hami-Kb}

\begin{figure}[h]
\vspace{5mm}
\begin{center}
\includegraphics[scale=0.65]{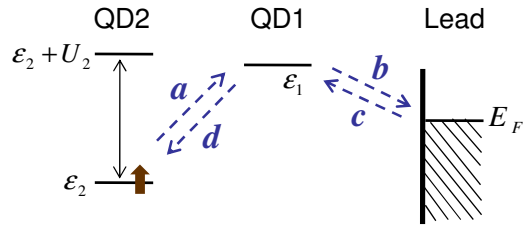}
\end{center}
\caption{
(Color online)  
Schematic energy diagram of a side-coupled DQD
for $\varepsilon_{1} >0$. This situation
is the same as that for Fig. \ref{sketchTDQD}(b).
In this figure, we describe the virtual tunneling events
by the (blue) dashed arrows, which are labeled by
$a, b, c$ and $d$.
The bold arrow at the QD2 indicates the local spin.}
\label{sketchprocess}
\end{figure}

We outline the derivation of the effective Hamiltonian
for the SE Kondo coupling given in
 Eq.\ \eqref{Hami-Kb} by using the perturbation theory 
in the tunneling matrix elements. \cite{Hewson-book}
The unperturbed ground state is chosen to be the one 
for $t=V_{s}=0$ and $-\varepsilon_2,\, \varepsilon_2+U_2,\, 
\varepsilon_1 \gg E_F\,(\equiv 0)$.
In this situation, the ground state of 
the two-impurity Anderson model defined in Eq.\ \eqref{Hami-TIAM}
is described by 
the singly occupied state at the QD2 and the empty one at the QD1.
We thus choose the unperturbed Hamiltonian to be
\begin{eqnarray} 
H_0 = H_{QD1} + H_{QD2} + H_{s}.
\label{unperturbH}
\end{eqnarray}
The unperturbed ground state is described by 
$d_{2\sigma}^\dag |F\rangle$, where $|F\rangle$ is the Fermi sea 
of the conduction band.
The tunneling terms 
$H_{T,s}$ and $H_{\rm int}$ are taken to be as the perturbation Hamiltonian, 
as illustrated in Fig.\ \ref{sketchprocess}.

The superexchange mechanism via the energy 
level $\varepsilon_1$ at the QD1 is described by  
 the fourth order perturbation with respect to the tunneling 
matrix elements $t$ and $V_{s}$.
In the fourth order perturbation, there are six processes.
Using the label for virtual electron tunnelings shown in 
Fig.\ \ref{sketchprocess}, these six processes are given by
\begin{align}
({\rm i})  & \quad a \to b \to c \to d &\qquad 
({\rm iv}) & \quad c \to d \to a \to b \nonumber\\
({\rm ii}) & \quad a \to c \to b \to d&\qquad
({\rm v}) & \quad c \to a \to b \to d \nonumber\\
({\rm iii}) & \quad a \to c \to d \to b &\qquad
({\rm vi})  & \quad c \to a \to d \to b . \nonumber
\end{align}
The virtual electron tunnelings described by the
labels $a$ and $d$ are caused by $H_{\rm int}$,
whereas those described by labels $b$ and $c$ 
are caused by $H_{T,s}$.
Therefore, the fourth order perturbation takes the form
\begin{eqnarray} 
H'_{1} \frac{1}{E-H_0} H'_{2} \frac{1}{E-H_0}
 H'_{3} \frac{1}{E-H_0} H'_{4},
\label{perturb}
\end{eqnarray}
where $H'_{j}$ for $j \in \{1,2,3,4 \}$ 
depends on the virtual processes,
namely $H'_j=H_{\rm int}$ for the virtual electron tunnelings labeled
$a$ and $d$, whereas $H'_j=H_{T,s}$ for those labeled $b$ and $c$.
As a representation of Eq.\ \eqref{perturb}  
in the subspace of $d_{2\sigma}^\dag |F\rangle$,
 we obtain the effective Hamiltonian in the form
\begin{eqnarray} 
H_{\rm eff}^{\lambda}
=
W^{\lambda} \sum_{k,k'}\sum_{\sigma} s_{k\sigma}^\dag s_{k'\sigma}^{}
+
J^{\lambda} \sum_{k,k'}\sum_{\sigma,\sigma'} 
s^{\dag}_{k\sigma} \frac{\vec{\tau}_{\sigma \sigma'}}{2} s^{}_{k'\sigma'} \cdot \vec{S}_2,\nonumber\\
\label{Heff-each}
\end{eqnarray}
where $\lambda$  runs over $({\rm i})$ to $({\rm vi})$. 
The couplings $W^{\lambda}$ and 
$J^{\lambda}$ are given by
\begin{eqnarray}
&&W^{({\rm i})}\!=\! -\frac{(V_s t)^2}
{2\varepsilon_2 (\varepsilon_2-\varepsilon_1)^2}, \quad
J^{({\rm i})}\!=\!4W^{({\rm i})},
\\
&&W^{({\rm ii})}\!=\! \frac{(V_s t)^2}
{2(\varepsilon_2-2\varepsilon_1)(\varepsilon_2-\varepsilon_1)^2}, 
\,\, J^{({\rm ii})}\!=\!-4W^{({\rm ii})},
\\
&&W^{({\rm iii})} \!=\! -\frac{(V_s t)^2}
{2\varepsilon_1 (\varepsilon_2-2\varepsilon_1)(\varepsilon_2-\varepsilon_1)}, 
\,\, J^{({\rm iii})}\!=\!-4W^{({\rm iii})},
\nonumber\\
&& \\
&&W^{({\rm iv})}\!=\! -\frac{(V_s t)^2}
{2(\varepsilon_1)^2 (\varepsilon_2+U_2)}, \quad
J^{({\rm iv})}\!=\!-4W^{({\rm iv})},
\\
&&W^{({\rm v})}\!=\! W^{({\rm iii})}, 
\quad J^{({\rm v})}\!=\!J^{({\rm iii})},
\\
&&W^{({\rm vi})}\!=\! \frac{(V_s t)^2}
{2(\varepsilon_2-2\varepsilon_1)(\varepsilon_1)^2}, 
\quad J^{({\rm vi})}\!=\!-4W^{({\rm vi})}.
\label{WJ-each}
\end{eqnarray}
Summing up these six processes, we 
obtain the effective Hamiltonian for the SE Kondo coupling, 
corresponding to $\mathcal{H}_{\rm K}^{\rm (b)}$ given in Eq.\ \eqref{Hami-Kb}.

For the finite Coulomb interaction $U_1$ at the QD1,
the processes of (ii), (iii), (v), and (vi) 
depend on $U_1$ since the QD1 is doubly occupied in  
the intermediate state. 
Thus for finite $U_1$,  $W^{\lambda}$ takes the form
\begin{eqnarray}
W^{({\rm ii})}\!\!&=&\!\! \frac{(V_s t)^2}
{2(\varepsilon_2-2\varepsilon_1-U_1)(\varepsilon_2-\varepsilon_1)^2}, 
\\
W^{({\rm iii})}\!\!&=&\!\! -\frac{(V_s t)^2}
{2\varepsilon_1(\varepsilon_2-2\varepsilon_1-U_1)(\varepsilon_2-\varepsilon_1)},\\
W^{({\rm v})}\!\!&=&\!\! W^{({\rm iii})},
\\
W^{({\rm vi})}\!\!&=&\!\! \frac{(V_s t)^2}
{2(\varepsilon_2-2\varepsilon_1-U_1)(\varepsilon_1)^2}. 
\label{W-each-finiteU1}
\end{eqnarray}
Correspondingly, $J^{\lambda}$ for 
$\lambda=({\rm ii})$, (iii), (v), and (vi)
is given by $J^{\lambda}=-4W^{\lambda}$, 
and thus we can obtain the exchange coupling $J_{\rm SE}$ 
given in Eq.\ \eqref{J-SE-U1}.



\end{document}